# A Hertzian Plasmonic Nanodimer as an Efficient Optical Nanoantenna

Andrea Alù, and Nader Engheta[*]

Department of Electrical and Systems Engineering, University of Pennsylvania

Philadelphia, PA 19104, U.S.A.

E-mail: andreaal, engheta @ee.upenn.edu

*Inspired by the geometry and shape of the classical radio-frequency radiator, the Hertzian dipole, here we analyze the design of a plasmonic optical dimer nanoantenna. We show how it may be possible to operate a pair of closely spaced spherical nanoparticles as an efficient optical nanoradiator, and how its tuning and matching properties may be tailored with great degree of freedom by designing suitable nanoloads placed at the dimer's gap. In this sense, we successfully apply nanocircuit concepts to model the loading nanoparticles. High levels of optical radiation efficiency are achieved, even considering the realistic absorption of optical metals, thanks to this specific geometry and design.*

## 1. Introduction

The pioneering work of Hertz at the end of the nineteenth century [1]-[3] is at the foundation of the modern antenna science and engineering, and therefore of an important part of nowadays wireless technology. His intuition of driving oscillating charges distributed over two closely spaced spherical capacitors (Fig. 1a) has proven successful for generating the first class of working radiators, and it has paved the way to myriads of wireless applications in the current technology. Nowadays, the theory and practice of RF antenna design is well established, and the old geometry of Hertz's first antennas (Fig. 1a) would definitely look outdated, compared with the millions of different antenna designs currently available for the numerous different purposes and applications.

---

[*] To whom correspondence should be addressed. E-mail: engheta@ee.upenn.edu

On a much different scale, recent advances in nanotechnology have led to the possibility of realizing and tailoring with high precision the geometry of clusters of nanoparticles up to a size of few nanometers. In particular, in several recent papers [4]-[15] it has been suggested how nanoparticles of different geometries may be employed as nanoradiators at optical frequencies. Bringing the antenna concepts up to the visible regime may indeed revolutionize wireless technology and communications, in terms of size reduction, speed and bandwidths of operation. However, for different reasons the optical nanoantenna science is still in its early stage, and the recent experiments on optical nanoantennas may be well compared with the first attempts performed by Hertz.

In this context, we have recently proposed a general theory that may bring and utilize the concepts of input impedance, radiation resistance, antenna loading and matching of optical nanoantennas, in order to translate the well known and established concepts of RF antenna design into the visible regime [15]. In particular, we have applied these concepts to the design and operation of nanodipoles [14]-[15], consisting of a pair of closely spaced thin nanorods, a geometry that may resemble one of the most commonly used RF antenna, the dipole. However, for various reasons nanodipole antennas provide a relatively lower radiation efficiency than classic RF dipoles [11], [14], and their realization might still be challenging from a technological point of view, due to the relatively high aspect ratio required for the nanostructures of which they are composed.

For these reasons, inspired by the pioneering work of Hertz, here we analyze in detail the radiation and loading properties of a plasmonic nanodimer in the form of two closely spaced spherical nanoparticles (Fig. 1b), which, with obvious differences in scale, may resemble the shape of Fig. 1a. To our initial surprise, the radiation properties of this class of optical nanoantennas prove to be far superior to those of nanodipoles of similar size, and their technological realization may become relatively less challenging, as different experimental groups have recently shown [16]-[18]. In the following, we report our theoretical and numerical results in the problem of nanodimer nanoantennas made of optical metals, showing the main physical mechanisms behind their high radiation efficiency and applying the nanocircuit paradigm [19]-[20] to study the loading and matching issues associated with this geometry. We believe that these findings may be readily verifiable experimentally within the limits of current nanotechnology and they may become of major interest for a wide range of applications, from optical communications to medical, biological and chemical sensing.

**2. Radiation Properties of the Nanodimer**

The nanoantenna geometry under analysis is depicted in Fig. 1b, and it consists of a plasmonic dimer of total length $L$ composed of two spherical nanoparticles separated by a gap of length $g \ll L$. Even if the total size of the nanoantenna is supposed to be of a few tens of nanometers, its size remains comparable with the optical wavelength of operation. This makes the nanodimer operation quite different from the RF Hertzian antenna of Fig. 1a, which was essentially meant to operate as a short dipole, due to its small electrical length (the operating wavelength in Hertz's first experiments was several meters long [2]). It is interesting to note, however, that the specific shape of the nanodimer (Fig. 1b), inspired with the Hertzian antenna, plays an important role in keeping its radiation efficiency high, as will be shown in the following, in a similar fashion as the spherical capacitors of the Hertzian antenna (Fig. 1a) have been proven to be essential in supporting the radiation from the alternating current driven along the dipole.

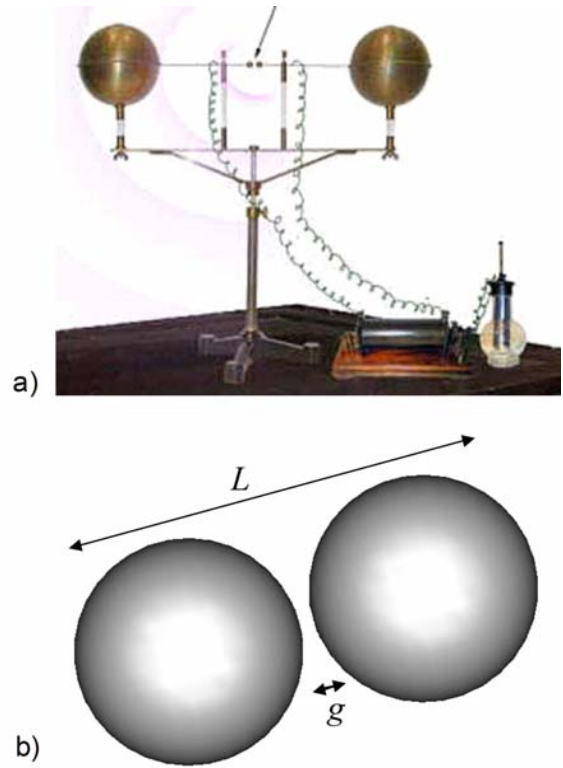

Figure 1 – (Color online). Analogy between two dimer antennas: (a) an early Hertzian antenna to operate at microwave frequencies (sketch from [3] http://www.sparkmuseum.com, reprinted with permission); (b) the nanodimer antenna under analysis here.

Figure 2 reports the calculated frequency dispersion for the peak of radiated electric field as a function of frequency for the geometry of Fig. 1b with $L = 100\,nm$, $g = 3\,nm$, when excited by a port with the internal resistance of $10\,k\Omega$. The port may model a feeding optical nanotransmission-line [21], an emitting quantum dot or a focused laser beam feeding the nanoantenna. The nanodimer is assumed to be made of silver, considering realistic frequency dispersion and losses for its permittivity. At the first (dominant) resonance, due to its limited electrical size, the nanoantenna radiates a donut-shape dipolar beam centered around the dimer axis. As seen from Fig. 2, the resonant peak is significantly shifted when different loading nanoparticles are considered at the gap. Following [14]-[15], we have considered three different insulating nanoloads, consisting of air ($\varepsilon = \varepsilon_0$), $Si_3N_4$ ($\varepsilon = 4.1\varepsilon_0$) and $Si$ ($\varepsilon = 13.37\varepsilon_0$) circular nanocylinders filling the dimer gap, all with circular cross section of radius $r = 3 nm$. A sketch of the loading geometry is reported in the inset of Fig. 2.

Following our work on optical nanocircuit elements [19]-[20], the interaction of these nanoparticles with the impinging light may be directly interpreted as a capacitive load (also the use of inductive loads may be envisioned by using plasmonic materials [14]-[15], [19]-[20] for the nanoloads). Consistent with our findings relative to the nanodipole geometry [14], an increase in the load capacitance, by choosing a loading material with larger permittivity, may significantly shift down the radiation resonance of the nanodimer. Two important features may be underlined in Fig. 2: the large variation of resonant frequency that may be achieved by simply varying the permittivity of loading nanoparticles at the gap, which spans several hundreds of $THz$, and the associated large sensitivity of the gap load. Both these features are clearly related to the specific shape of this nanoantenna geometry, which ensures a large concentration and specific orientation of the electric field at the gap location, interacting with the nanoloads.

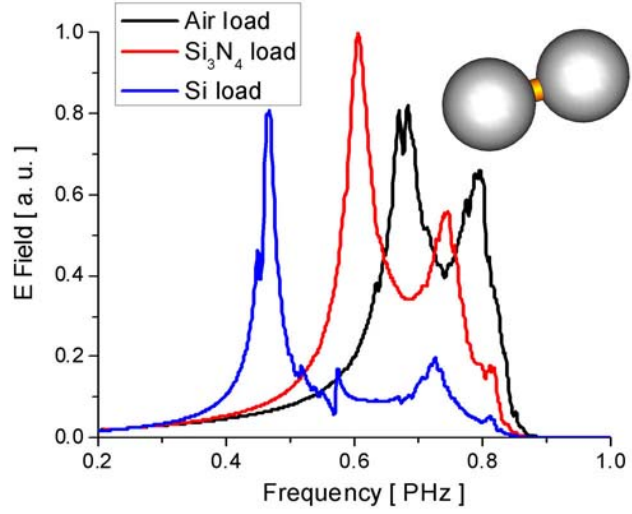

Figure 2 – (Color online). Peak in the radiated electric field versus frequency for the geometry of Fig. 1b with $L = 100\,nm$, $g = 3\,nm$ and $r = 3\,nm$, fed by a $10\,k\Omega$ port.

Figure 3 reports the extracted input impedance $Z_{in}$ evaluated at the gap of the loaded nanoantennas of Fig. 2, versus the load permittivity. With the thin magenta line we have also reported the input impedance of the "unloaded" (or "de-embedded" or "intrinsic") dimer $Z_{dimer}$, which is obtained from any of the loaded (thick) curves after de-embedding the parallel capacitance represented by the corresponding load. This is consistent with the circuit model for the loaded nanodimer reported in the inset of Fig. 3, for which the intrinsic impedance of the nanoantenna $Z_{dimer}$ is in parallel with the nanoload impedance [14]-[15], [19] given by:

$$Z_{load} = \left( -i\omega \varepsilon_{load} \frac{\pi r^2}{g} \right)^{-1}. \tag{1}$$

This value is easily de-embedded after post-processing in order to evaluate the "intrinsic" input impedance of the unloaded dimer $Z_{dimer}$. It is interesting to note that, even though the three dashed lines in Fig. 3 have been independently extracted from full-wave numerical simulations (using CST Microwave Studio™ [22]) of the different loaded geometries, they all provide the same unloaded input impedance $Z_{dimer}$, reported as the thin line in Fig. 3, after proper de-embedding of the corresponding value (1) for $Z_{load}$. This is possible because of the specific orientation of the electric field in the gap region, parallel to the dimer axis, which ensures the functionality of the nanoload as lumped nanocapacitors, as described by (1).

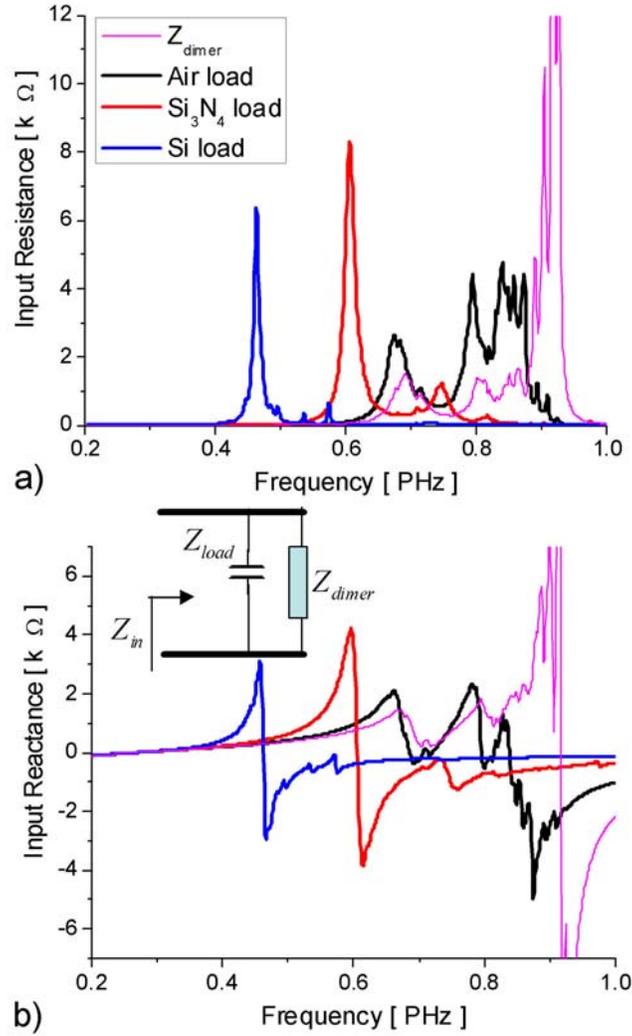

Figure 3 – (Color online). Input impedance $Z_{in}$ at the port gap for the geometries of Fig. 2 varying the load permittivity (thick lines) and corresponding extraction of $Z_{dimer}$ (thin).

Figure 2 and 3 are indeed very informative: due to the large internal impedance of the port that we have considered, the maximum radiation peak in Fig. 2 is achieved near the "open-circuit" resonance of $Z_{in}$ (following the terminology of [14]), i.e., in the range of frequencies for which the parallel between $Z_{dimer}$ and $Z_{load}$ enter into resonance, producing high real values of $Z_{in}$. As evident from Fig. 3, a proper design of $Z_{load}$, given by (1), may suitably shift the radiation peak over a wide frequency range at will. This implies that the tuning and matching of a nanodimer antenna may be relatively easily achievable by slightly varying the geometrical or electromagnetic parameters of a nanoparticle placed at its gap. This large tunability is associated with the specific geometry of the nanodimer.

Figure 4 reports the electric and magnetic field distributions on the E plane (snapshots in time) for the nanodimer of Fig. 2-3 with an air load (i.e., empty gap) at the resonant frequency $f = 665 THz$. A strong concentration of the electric field in the tiny gap region is evident in the plots, as expected due to the specific nanodimer geometry. This makes the antenna operation sensitive to a change in the permittivity of the gap region, and it explains the strong resonances associated with the input impedance extracted at the gap, which may be properly shifted and tuned by varying the load permittivity using a variety of method such as electronically, optically or magnetically.

It is also evident in Fig. 4a that the displacement current injected by the feeding port enters the metallic nanodimer (due to its poor conductive properties at these high frequencies [14]) and it may rapidly "spread" all over the nanodimer volume. This spreading of optical displacement current makes the nanodimer operation significantly different from the nanodipole [14], whose small nanorod diameter confines the displacement current in a limited space. This different behavior results in a significantly larger robustness of the nanodimer to material losses, as we discuss in the following. The reduced concentration of displacement current allows also shifting up the resonant frequency of this nanoantenna, compared to a nanodipole of same length $L$. For instance, a cylindrical nanodipole of same length $L=100\,nm$ and radius $r=5\,nm$ would resonate around $350\,THz$ (almost twice the optical wavelength) [12], [14]. This phenomenon is very counterintuitive if viewed in terms of classic RF design, since a reduction of the antenna physical volume should have intuitively corresponded to an increase in the resonance frequency. However, as already noticed in several other plasmonic resonant setups, an increase in the volume of a plasmonic nanodevice may surprisingly shift up its resonant frequency of operation, due to less concentration of the fields around the plasmonic interfaces [13]-[14], [21], [23]. This is exactly the reason for this different resonant behavior, as confirmed by the spreading of displacement current in Fig. 4a. The magnetic field distribution in Fig. 4b confirms the dipolar resonant characteristics of the nanodimer antenna.

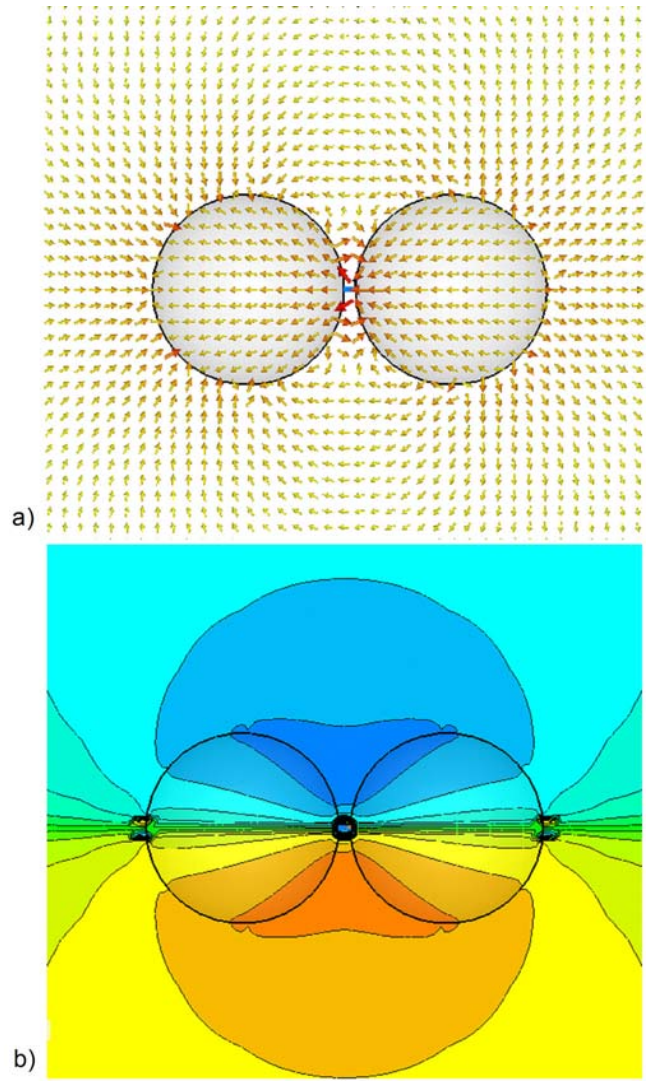

Figure 4 – (Color online) Electric (a) and magnetic (b) field distributions (snapshots in time) on the E plane of the nanodimer antenna fed at its gap at the resonance frequency $f = 665 THz$.

Figure 5a reports the calculated antenna gain $G$ vs. frequency (as commonly done for RF antennas, here the gain does not consider the port mismatch, but considers the power effectively entering the antenna). It is evident that the gain can reach values comparable with RF dipolar antennas over a broad range of frequencies centered around the resonant frequency. For this geometry, the maximum gain is achieved around $700 THz$, which coincides with the resonance frequency of the air-loaded nanoantenna. However, good performance may be achieved over a very wide frequency range.

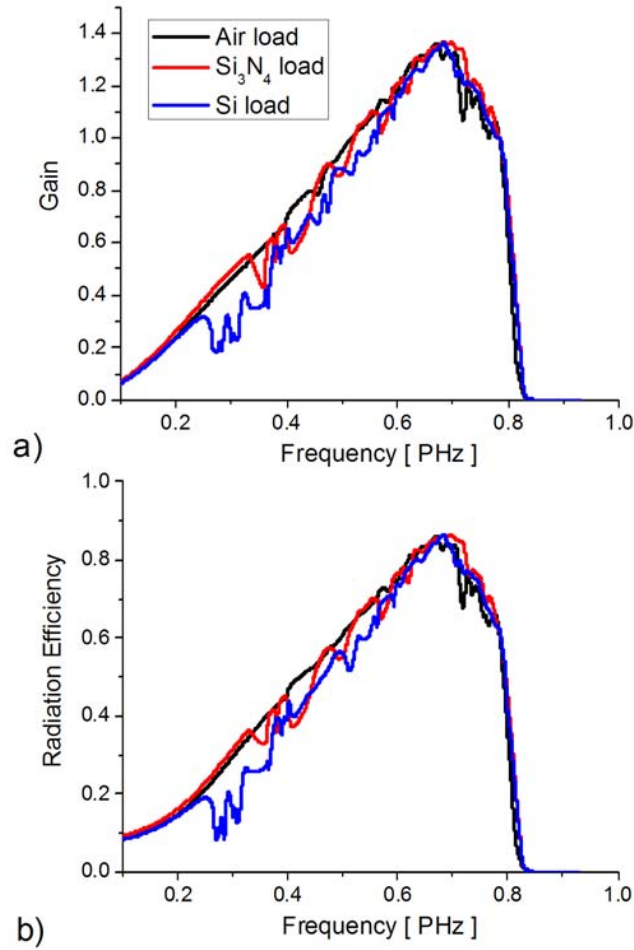

Figure 5 – (Color online) Gain (a) and radiation efficiency (b) as a function of frequency for the antennas of Fig. 2-3.

Since the nanodimers considered here are relatively short compared to the wavelength, their directivity is $D \simeq 1.5$ over the whole frequency range of interest. Their radiation efficiency $\eta_{rad} = G/D$, defined as the ratio between the radiated power and the power effectively entering the nanodimer volume, is easily evaluated in Fig. 5b. Interestingly, the radiation efficiency of the nanodimer may reach impressive values close to 90%, which is significantly higher than the radiation efficiency associated with optical nanodipole antennas [11], [14]. As anticipated above, this is related to the spreading of displacement current in each of the two nanoparticles composing the dimer. Compared to the constrained confinement of the guided mode along a thin and long nanodipole, this spreading allows a much more robust response to material absorption, which reflects in much higher radiation efficiency, as predicted in Fig. 5. This makes the nanodimer geometry particularly suitable for optical antenna operation.

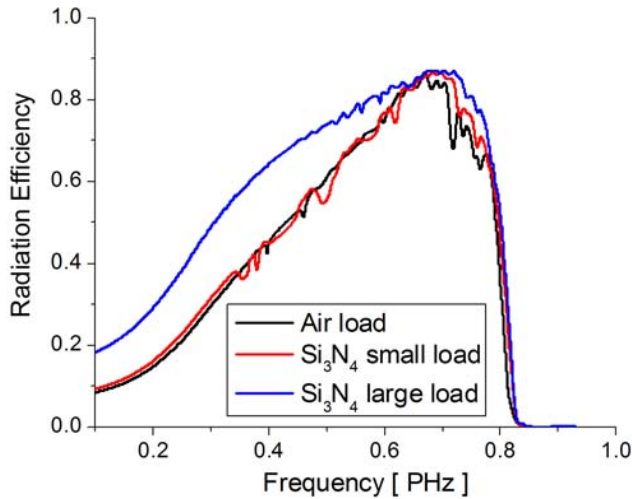

Figure 6 – (Color online) Consistent with Fig. 5, radiation efficiency as a function of frequency varying the size of the load from $r = 0$ (black line), $r = 3\,nm$ (red), to $r = 6\,nm$ (blue).

Figure 6, reports the radiation efficiency varying the $Si_3N_4$ load size (in particular its radius $r$) for the nanodimer of Figs. 2-3. The value of $\eta_{eff}$ is not sensibly affected by a change in the load geometry, providing high radiation efficiencies at the resonance frequency, independent of the specific geometry of the loading

nanoparticle. The variation in the load clearly affects, however, the resonance properties of the nanodimer, as discussed above.

## 3. Scattering Properties of the Nanodimer

By reciprocity, the scattering and receiving properties of the nanodimer when excited by an external excitation (plane wave or Gaussian beam) are expected to support similar resonance properties as highlighted in the previous section, together with a strong tunability by varying the nanoloads placed at the gap. Figure 7 reports the variation of the scattering cross section (SCS) of the nanodimer vs frequency with different permittivity of the nanoload at its gap, considering the geometries of Fig. 2-3.

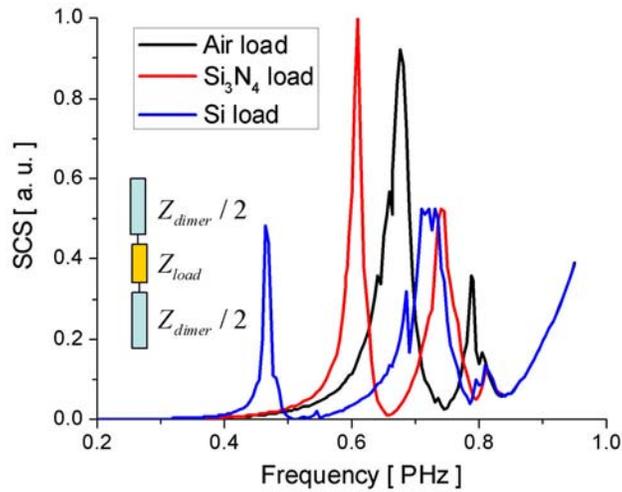

Figure 7 – (Color online) Scattering cross section for the nanodimers of Fig. 2-3 varying the permittivity of the nanoloads.

Consistent with our findings in the nanodipole geometry [15], the scattering problem may be modeled as in the circuit sketch in the inset of Fig. 7, using the same quantities evaluated in the previous section. This requires, as confirmed by our simulations in Fig. 7, that the resonant peaks in the SCS are located at the "open-circuit" resonances of the input impedance of the loaded nanodimer (Fig. 3, thick lines). Fig. 7 confirms that also the scattering and receiving properties of the nanodimer may be tuned at will by varying the nanoload permittivity. The variety of available optical materials with different permittivity ensures several degrees of freedom for tuning the nanodimer response at will over a wide frequency range that covers the optical and infrared range of frequencies. Moreover, the nanodimer geometry may ensure strong resonant scattering and radiation efficiency also in its receiving/scattering operation. With a similar technique as the one presented in [15], it may be possible to load the nanodimer antenna with parallel or series combinations of nanoparticles, which may provide an even greater tunability of these resonant properties.

## 4. Conclusions

In this work we have analyzed in details the radiation properties of an optical nanodimer. In particular, we have shown how its peculiar geometry may provide much larger radiation efficiency and resonance tunability than with optical nanodipoles, due to the spreading of displacement current that enters the nanodimer at its gap. We have also discussed how its radiation properties may be tuned at will by using the nanocircuit concepts to interpret its interaction with nanoparticles used as loads at its gap. Similarly, we have also shown how the nanoloads may help matching the input impedance of the antenna to a given feeding/receiving optical network by varying and tuning the optical input impedance at the feeding/receiving point (dimer gap). In this sense, it may be possible to operate efficient nanoantennas at optical frequencies using the familiar concepts of RF antennas transplanted at much higher frequencies. These findings may find useful applications in various disciplines spanning optics, medicine and biology, where the need for imaging, sensing and communications at the nanoscale and at a fast rate is relevant at present times.

**Acknowledgments**

This work is supported in part by the U.S. Air Force Office of Scientific Research (AFOSR) grant number FA9550-05-1-0442.